\documentclass[twocolumn, prb]{revtex4}

\usepackage{graphicx}
\usepackage{dcolumn}
\usepackage{bm}

\begin{document}

\title{ N\'eel and disordered phases of coupled 
Heisenberg chains with $S=\frac{1}{2}$ to $S=4$}  

\author {S. Moukouri }

\affiliation{ Department of Physics and  Michigan Center for 
          Theoretical Physics \\
         University of Michigan 2477 Randall Laboratory, Ann Arbor MI 48109}

\begin{abstract}
We use the two-step density-matrix renormalization group method to study the 
effects of frustration in Heisenberg models for $S=\frac{1}{2}$ to $S=4$ in 
a two-dimensional anisotropic lattice. We find that as in $S=\frac{1}{2}$ 
studied previously, the system is made of nearly disconnected chains at the 
maximally frustrated point, $J_d/J_{\perp}=0.5$, i.e., the transverse 
spin-spin correlations decay exponentially. This leads to the following 
consequences: (i) all half-integer spins systems are gapless, behaving like
 a sliding Luttinger liquid as in $S=\frac{1}{2}$; (ii) for integer spins, 
there is an intermediate disordered phase with a spin gap, with the width 
of the disordered state is roughly proportional to the 1D Haldane gap.
\end{abstract}

\maketitle

\section{Introduction}

 An important theorem by Dyson, Lieb, and Simon (DLS)  states 
 that the  Heisenberg Hamiltonian on bipartite lattices with $S \ge 1$ 
have long-range order in the ground state \cite{dls,neves}. We now
know from quantum Monte Carlo simulations \cite{young} that this result extends 
to $S=\frac{1}{2}$ systems. There is current
interest as to how a disordered state emerges out of the N\'eel state.
This question is important to the physics of frustrated materials
where some systems exhibit no magnetic order down to the experimentally
accessible temperatures and are thus believed to be disordered in the
ground state. A disordered phase may also be relevant in the theory
of high temperature superconductors. This issue and the related one of the
 eventual role of Berry phases was hotly debated in the
late 1980's and early 1990's soon after the discovery of the high $T_C$
materials \cite{dombre, stone, zee, affleck1, haldane2}. 
Many possible disordered states were proposed but none of
them gained consensus. For an extensive discussion on this topic, we refer
the reader to the book by Fradkin \cite{fradkin}.

In the search of the nature of a ground state of a quantum Hamiltonian,
there is another useful theorem by  Lieb, Schultz, and Mattis (LSM)\cite{lsm},  
initially formulated for 1D systems and later extended to 2D systems by 
Affleck \cite{affleck1}  restricts the possible ground 
states of half-integer spin systems. Either the ground state is degenerate, 
(presumably due to  a broken symmetry)  or it is unique and gapless without 
any long-range order. For Heisenberg Hamiltonians, the latter possibility 
appears difficult to realize  in 2D because dimer-dimer or spin-spin 
correlations have a power law decay in 1D. It would thus be expected 
that interchain couplings will lead to long-range order in one of the two
 channels. If the magnetic order is frustrated, a dimerization would be
expected. Frustrated spin models often involve competitions between 
two magnetic orders, the expected dimerized phase is believed to lie
between these magnetic phases. A possible alternative  to this scenario
allowed by the LSM theorem is the occurence of a disordered gapless  
state at transition between the magnetic phases.

In recent publications \cite{moukouri-TSDMRG, moukouri-TSDMRG2}, we have studied the possible emergence of a disordered state in a model of coupled 
Heisenberg chains. This model is a spatially
anisotropic version of the well studied $J_1-J_2$ model. The
model essentially retains the physics of the $J_1-J_2$ model: it presents 
a phase transition between two magnetic phases. The first phase is a
 N\'eel phase characterized by the ordering wave vector $Q=(\pi,\pi)$. The 
order parameter in this phase is maximal in absence of the diagonal coupling 
$J_d$. It decreases as $J_d$ is increased until it vanishes at the maximally 
frustrated point. Beyond this point, another N\'eel state with $Q=(\pi,0)$ 
becomes the ground state. The existence of these two phases was predicted in 
numerous studies \cite{lhuillier} of the isotropic ($J_{\perp}=1$) version of 
this model. But what has caused the continuing interest into this simple model 
is the question of whether there is an intermediate phase between these two 
magnetic phases. Simple physical arguments suggest the existence of such a 
phase because the two states are associated with subgroups of the  
 $SU(2)$ symmetry group of the spin Hamiltonian and of the $C_4$ symmetry 
group of the square lattice that neither include each 
other. From the Landau theory we know that a continuous transition from 
these two phases is forbidden. Initial suggestions for the possible 
intermediate phase was the resonating valence bond 
(RVB) \cite{baskaran} phase, flux phases \cite{marston, kotliar}, 
chiral spin liquid \cite{wen} or the spin-Peierls (SP) \cite{read}. 
 The SP phase has lately emerged as the front runner. It
causes the same type of difficulty as the direct transition between 
two magnetic phases. Since in that case, the transition  is between a magnetic 
state that breaks the spin rotational symmetry but not the lattice 
translational symmetry to a dimerized state that breaks the lattice 
translational symmetry but not the spin rotational symmetry. The transition
 to SP phase has triggered interesting proposal of an extension
 of the conventional Landau-Ginzburg-Wilson (LGW) theory of second order 
phase transition\cite{dc}.

It seems, however, that this theory does not apply to the $J_1-J_2$ model, 
where we did not find any evidence of an intermediate phase for 
$S=\frac{1}{2}$. Numerical
data suggests a transition, which seems to be of second order, between the 
two magnetic phases at the maximally
frustrated point for both the anisotropic and isotropic 
models \cite{moukouri-TSDMRG3}. At the transition 
point, the competing magnetic orders neutralize each other and the system 
behaves like a collection of loosely bound chains, even if the bare 
interactions are not small. Classically, the 
ground state is degenerate at this point. 
 This degeneracy is lifted by quantum 
fluctuations. In Ref.\cite{moukouri-TSDMRG2,moukouri-TSDMRG3}, we have 
shown that  among all the possible clusters, chains offer the best compromise 
between minimizing the energy and avoiding frustration at the same time. 
At the maximally frustrated point, the transverse interactions seem to
be irrelevant. I.e., up to the largest lattice size studied, transverse
spin-spin correlations decay exponentially and the longitudinal correlations
revert to those of decoupled chains. 
This disordered state is a singlet and gapless consistent with the LSM 
theorem\cite{lsm,affleck1}. It is reminiscent of a sliding
Luttinger liquid (SLL) found in models 
of coupled fermions chains \cite{carpentier,kivelson,kane}. It thus appears 
that the intermediate region where 
a disordered phase has long been thought to exist is just a critical region. 
That is the reason why it has resisted various approaches for nearly two 
decades. At the maximally frustated point, the correlation functions are 
1D-like, thus the rotational spin symmetry of the system is restored. 
In fact such a transition between these two magnetic phases already exists 
in the unfrustred model when the transverse exchange parameter, $J_{\perp}$, 
is varied from positive values to negative values. At the point $J_{\perp}=0$,
 there is a transition from 2D to 1D, where properties are identical to that 
of the maximally frustrated point. The only difference between the case 
$J_{\perp}=0$ and the maximally frustrated point is the presence of 
irrelevant transverse terms which do not change the long distance behavior 
of the correlation functions as shown in Ref.\cite{moukouri-TSDMRG2}. From this 
result the LGW theory applies if it is  assumed that the system's 
group of symmetry at the critical point contains the groups of 
symmetry of the two magnetic phases.

 It is important to study how this interesting physics extends to larger 
spin systems. First, because many frustrated systems contain larger spins. 
Second, because there are  some interesting predictions from 
large $S$ approaches
 about the emergence of a disordered phase from a N\'eel phase as function
of $S$. Affleck \cite{affleck1} argued that since the LSM theorem does not 
apply to integer spin systems, there might be a distinction between 
integer and half-integer spin systems in 2D as well. Haldane \cite{haldane2} discussed that in addition to the now 
well-established distinct behavior between half-integer and integer 
spins in one dimension \cite{haldane1}, there might be a difference between 
odd and even integer spins in two dimensions due to the effects of the 
Berry phase. Read and Sachdev carried out a large $N$ analysis of the possible
 disordered phase as function of the value of the large $N$ equivalent of
the spin. Their results  were consistent with Haldane's predictions. 
Three types of disordered states 
 were predicted. For half-integer spins, the non-magnetic phase 
is a  SP phase which breaks the translational symmetry along the two 
directions of the square lattice. For odd integer spins, the  
non-magnetic state is made of weakly-coupled chains, i.e., the translational 
symmetry is broken along one direction only. Finally for even integer spins, 
 the disordered state are valence bond solids, like the 
Affleck-Kennedy-Lieb-Tasaki (AKLT) state\cite{aklt}, i.e, it does not 
break any translational symmetry.

 In the large $S$ approaches the Heisenberg model is mapped onto the 
non-linear sigma model ($NL\sigma$) with a Berry phase term. This mapping
 is only approximate\cite{haldane1,affleck} and there can be some
subtle differences with the original model\cite{affleck}. Indeed, 
the realization of the Haldane conjecture in 1D shows their power. But, 
in absence of exact results for small $S$, it is impossible to know whether 
their predictions of a disordered phase extends to small $S$. Another potential 
 problem is that the mapping to the $\sigma$-model assumes the presence 
of a smooth configuration of spins. This is true in the weak-coupling regime 
(N\'eel ordered phase). But this assumption may break down in the strong 
coupling regime (disordered phase).
 In one dimension, the $\sigma$-model coupling constant is given by
 $g=2/(\sqrt{S(S+1)}\sqrt{1-4J_2})$ \cite{affleck}. Thus for $S=\frac{1}{2}$,  
the equivalence between the two models breaks down at $J_2=0.25$, i.e., 
close to the transition to a dimerized state.
 Such a breakdown seems to occur in the $J_1-J_2$ model where, as seen 
above, the large $N$  predictions conflict for $S=\frac{1}{2}$ with the 
TSDMRG in the $J_1-J_2$ model.   But since spin-half integer systems are 
critical, it could be objected that the behavior seen in our numerical 
studies are 
due to finite size effects. For large enough lattices there might be a relevant 
interaction which can drive the system to a SP phase as predicted from 
large $N$. Though this scenario appears to be unlikely, as discussed in 
Ref.\cite{moukouri-TSDMRG3}, it cannot be completely rejected. In principle,
 one would expect a different behavior for integer spin systems which are 
known to have a spin gap in 1D \cite{haldane1}.

In this paper, we applied the TSDMRG to study 
$S=\frac{1}{2}$, $1$, $\frac{3}{2}$,...,$4$
systems in the anisotropic 2D Hamiltonian (\ref{hamiltonian}). Our results
in the absence of frustration are in agreement with the DLS theorem 
Ref.\cite{dls,neves}. We find that for all $S$ the ground state is ordered 
in the absence of frustration. For all $S$ except for $S=\frac{1}{2}$, the 
order parameter is large enough so that the extrapolated value are reliable. 
This result constitutes a non-trivial test of the TSDMRG, since the TSDMRG 
starts from decoupled chains which are disordered. When the frustration is 
turned on, the general mechanism found for the destruction of the N\'eel phase 
is the severing of the frustrated bonds in the transverse direction, leading 
to a disordered state with the transverse correlations that decay 
exponentially at the critical point as previously found for $S=\frac{1}{2}$. 
However, a different conclusion is to be drawn for half-integer $S$ and for 
integer $S$ for which the LSM theorem does not apply. All half-integer systems 
are similar to $S=\frac{1}{2}$. The disordered state is confined at the 
critical point, it has a SLL character. But for integer $S$, because of the 
Haldane gap $\Delta_H$ in the chain, there is an intermediate phase whose width is roughly $ \propto \Delta_H$.

This paper is organized as follows. In the next section 
we discuss the model and the method.  In section 
(\ref{spin1}), we present extensive results for $S=1$ systems. This analysis is
similar to the one made for spin $\frac{1}{2}$ systems in 
Ref.\cite{moukouri-TSDMRG3}. In section (\ref{spinS}) the results
for systems with $S=\frac{1}{2}$ to $4$ are presented. 
In section (\ref{conclusions}), we present our conclusions.

\section{Model and Method}

\subsection{model} 

We apply the TSDMRG \cite{moukouri-TSDMRG,moukouri-TSDMRG2} to the 
spatially anisotropic Heisenberg Hamiltonian,

\begin{eqnarray}
 \nonumber H=J_{\parallel} \sum_{i,l}{\bf S}_{i,l}{\bf S}_{i+1,l}+J_{\perp} \sum_{i,l}{\bf S}_{i,l}{\bf S}_{i,l+1}\\
+J_d \sum_{i,l}({\bf S}_{i,l}{\bf S}_{i+1,l+1}+{\bf S}_{i+1,l}{\bf S}_{i,l+1})
\label{hamiltonian}
\end{eqnarray}

\noindent where $J_{\parallel}$ is the intra-chain exchange parameter and is set
 to 1; $J_{\perp}$ and $J_d$ are respectively the transverse and diagonal 
interchain exchanges. This model is the object of current interest 
\cite{tsvelik, moukouri-TSDMRG2, sindzingre, starykh}. It is
a starting point of understanding the $J_1-J_2$ model which is recovered
when $J_{\parallel}=J_{\perp}=J_1$ and $J_d=J_2$. It retains the basic
physics of the $J_1-J_2$ model and has the advantage that  in the
limit $J_{\perp},J_d \ll 1$, well tested 1D results can be used to initialize
a perturbative RG analysis. 
 
\subsection{The two-step DMRG}

In the TSDMRG, to study a 2D lattice of size $L \times (L+1)$ ( we will 
refer to the 2D
systems only by their linear dimension $L$), we start by applying the 
usual 1D DMRG ($m_1$ states are kept) or exact diagonalization (ED) 
to a single chain $l$ of length $L$ to obtain $m_2$ low lying eigenstates 
and eigenvalues, $\phi_{n_l}$, $\epsilon_{n_l}$, $n_l=1,2,...m_2$, 
respectively. Then, we formally write the tensor product of the $L+1$ 
chains,

\begin{equation}
\Phi_{[n]}= \phi_{n_1} \phi_{n_2} ...  \phi_{n_{L+1}}.
\end{equation}

$\Phi_{[n]}$ is an eigenstate of the Hamiltonian with $J_{\perp}=0.$ and
$J_d=0$, $[n]=(n_1,n_2,...,n_{L+1})$. The $\Phi_{[n]}$ constitute a 
many-body basis of the truncated Hilbert space of the tensor product of
$L+1$ chains. The corresponding eigenvalue 
is:

\begin{equation}
E_{[n]}= \epsilon_{n_1}+\epsilon_{n_2}+...+\epsilon_{n_{L+1}}
\end{equation}
 
 The 2D Hamiltonian (\ref{hamiltonian}) is then projected onto this truncated 
basis to yield an effective one dimensional Hamiltonian which is studied 
using the DMRG. 

The TSDMRG is  perturbative; but the expansion is made onto the 
smaller term of the Hamiltonian itself not on the Green's function or the 
ground-state wave function. We have shown that starting from a disordered 
state, the TSDMRG is able to reach the ordered state without any addition 
of a term that explicitly breaks the symmetry such as a magnetic field.
The TSDMRG was tested against the quantum Monte Carlo (QMC) method in 
Ref.(\cite{moukouri-TSDMRG2}) and against ED in Ref.(\cite{alvarez}). 
The TSDMRG is variational, its performance can systematically be improved
by increasing $m_1$ and $m_2$.  Key indicators about the performance of
the TSDMRG are the truncation error $\rho_1$ during the first step, the width 
$\delta E$ of the $m_2$ states kept and the truncation error $\rho_2$ during 
the second step. In principle, it is necessary that the ratio of $\delta E$ 
over the transverse coupling be large for the TSDMRG to yield great accuracy.
Typically, one must have $\delta E /J_{\perp} \approx 10$. If this condition
is fulfilled and the $m_2$ states are accurate enough, .ie., $\rho_1$ is
small, the TSDMRG method can reach the QMC accuracy. So far, this has
been achieved only for small couplings $J_{\perp} \alt 0.1$ and lattice
sizes of up to $L=16$ keeping up to $m_2=96$. The amount of calculations
involved remains modest and so far are done on a workstation. The accuracy 
decreases by increasing $J_{\perp}$ leading to less accurate results in the
ordered state. But when both $J_{\perp}$ and $J_d$ are turned on, the 
performance of the TSDMRG becomes more complex as we will see below.

\begin{figure}
\includegraphics[width=3. in, height=2. in]{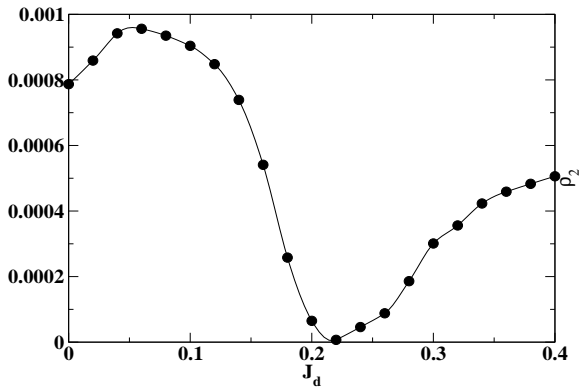}
\caption{Truncation error in the second TSDMRG step as function of $J_d$ for
$J_{\perp}=0.4$, $S=1$ and $L=10$, ED was performed in the first step and in
the second step, $m_2=64$.}
\vspace{0.5cm}
\label{err}
\end{figure}

In this work, the calculations for $S=1$ were performed similarly to those 
for spin $S=\frac{1}{2}$ systems in Ref.\cite{moukouri-TSDMRG3}. In most cases, 
ED was applied during the first step. In some case, i.e, some runs of $L=10$ 
and all runs of $L=12$, a single DMRG iteration was used. For instance 
for $L=10$ and $S=1$, when $m_1=81$ states are kept, a single $DMRG$ iteration 
is necessary to reach desired size. For this calculation, we kept up to 
$m_2=96$ 
states during the second TSDMRG step. For the maximum performance of the 
algorithm, it is necessary that $J_{\perp}$ be in the order of the finite 
size gap of the single chain $\Delta_L$. For $S=1$, 
$\Delta_L \rightarrow \Delta_H$ when $L \rightarrow \infty$, where
$\Delta_H =0.4107$ \cite{white} is the Haldane gap.
 A second condition to fulfill is $\delta_L \gg J_{\perp}$ where 
$\delta_L$ is the width of the retained eigenvalues.  
 As noticed in Ref.\cite{moukouri-TSDMRG3},
the TSDMRG is more accurate in the highly frustrated regime. In 
Fig.(\ref{err}) we show the truncation error, $\rho_2$, when two states 
are targeted in the second step as function of $J_d$ for $S=1$ and 
$J_{\perp}=0.4$. $\rho_2$ is minimal near $J_d=0.22$.
 At this point the system is an assembly
of nearly disconnected chains; the DMRG is thus expected to perform better.

\subsection{Illustration in the $S=1$ case}

We now wish to provide a detailed description of a typical TSDMRG
calculation. For this illustration, $S=1$ and  $L=12$, i.e, following 
the convention set above, the size of the 2D lattice is $12 \times 13$.
We start the usual 1D DMRG iteration keeping $m_1=243$ 
states, i.e, the initial superblock size is $L=12$. At this point the
DMRG is equivalent to ED. At the next iteration, the superblock size
is $L=14$, its total number of states is $M_s=(3 \times 243)^2$. 
The size of the reduced superblock (the superblock minus
the two single site blocks) size is $L=12$. During this iteration, 
spins sectors with $S_T^z=0,\pm 1, \pm 2, \pm 3$ are targeted. The 
lowest states in each of these sectors for $L=12$ have respectively the
following energies, $E_0=-16.8696$, $E_{\pm 1}=-16.3854$, 
$E_{\pm 3}=-16.2780$. The truncation error is $7.3 \times 10^{-7}$.
The reduced superblock is then diagonalized and the $m_2=64$ lowest
lying states are kept. The energy of the highest state among these
$m_2$ states is $-14.2687$ which is lower than $E_{\pm 4}=-12.6524$,
the lowest states of the $S_T^z=\pm 4$ sectors. For this reason, these sectors
was not targeted. The operators ${\bf S}_i$ at each site are stored and
updated.

From the energy levels above, we see that the 
finite size spin gap is $\Delta_L=0.4842$, which is not very far
from its value $\Delta_H=0.4107$ in the thermodynamic limit. Our
choice of $J_{\perp}=0.4$ ensures that the chains will effectively
be coupled at this size. The matrix $O$ whose columns are made of
the $m_2$ vectors $\phi_n$ kept is used to express all the operators in the
truncated reduced superblock basis, 

\begin{equation}
 {\bf {\bar S}}_i=O^{\dagger}{\bf S}_i \delta_{3,3'}O,
\label{updates}
\end{equation}   

\noindent the intra-chain Hamiltonian is likelywise updated 

\begin{equation}
h=O^{\dagger} (h_{B_1}\delta_{3,3'}+h_{B_3}\delta_{1,1'})O.
\label{updateh}
\end{equation}

\noindent In these equations, we have adopted the usual convention
that the different blocks of the superblock are labelled $1-2-3-4$. 
For PBC, blocks $2$ and $4$ are made of a single site
and blocks $1$ and $3$ are the largest blocks. In Eq. (\ref{updates}),
it is supposed that the spin to update is in block $1$. In Eq. (\ref{updateh}),
$h_{B_i}$ represents the internal Hamiltonian of block $i$. The first step 
ends with the updating of these operators.

Each chain may now be viewed as a super 'spin'
with additional internal degrees of freedom due to the different sites.
A chain $l$ is described by its 'spin' value 
${\tilde {\bf S}}_l=({\bf {\bar S}}_{il}, i=1,...,L)$ and its internal 
Hamiltonian
$h_l$. $h_l$ is diagonal in the basis of the $m_2$ states kept. The
effective first order Hamiltonian which approximates the original 2D
Hamiltonian is now given by

\begin{equation}
H_{eff}=\sum_l h_l + J_{\perp} \sum_l {\tilde {\bf S}}_l * {\tilde {\bf S}}_{l+1}
          + J_d \sum_l {\tilde {\bf S}}_l \times {\tilde {\bf S}}_{l+1},
\end{equation}

\noindent where,

\begin{equation}
{\tilde {\bf S}}_l * {\tilde {\bf S}}_{l+1}=\sum_i {\bf {\bar S}}_{i,l} {\bf {\bar S}}_{i,l+1}
\end{equation}

\noindent and

\begin{equation}
{\tilde {\bf S}}_l \times {\tilde {\bf S}}_{l+1}=\sum_i {\bf {\bar S}}_{i,l} 
{\bf {\bar S}}_{i+1,l+1}+{\bf {\bar S}}_{i+1,l}{\bf {\bar S}}_{i,l+1}.
\end{equation}

We then proceed to compute the low lying states of $H_{eff}$ using the
conventional DMRG again. For this simulation we keep $m_2$ states and use
$3$ blocks instead $4$ to form the superblock.

As expected from the study of $S=\frac{1}{2}$ systems, the TSDMRG is more
accurate in the highly frustrated regime than in the unfrustrated case.
 $\rho_2$ is relatively  large in the unfrustrated regime because of the
relatively large value of the interchain coupling. The same simulation
with $J_{\perp}=0.2$ leads to an improvment of factor $10$. It is worth
noting that the superblock size in 
this step is $m_2^3$, we are able to reach $m_2=100$ on a workstation, 
this remains modest with respect to what can be achieved on today's 
supercomputers. For the multi-chain DMRG superblock sizes of about $100$ 
times larger are accessible \cite{jeckelmann}, this means that it is possible
to reach $m_2 \approx 500$ on a supercomputer. This would increase the current 
accuracy of the TSDMRG by two or more orders of magnitude. 
This shows the great potential of the TSDMRG. These possibilities are 
under exploration.

\section{Results for $S=1$}
\label{spin1}

The case of spin $\frac{1}{2}$ has been extensively studied in 
Ref.\cite{moukouri-TSDMRG2,moukouri-TSDMRG3}. We were able to reach lattice 
sizes of up to $64 \times 65$ and show that as seen in QMC simulations, 
 the system is ordered in absence of frustration for small $J_{\perp}$. 
But when $J_d \neq 0$, we have shown that in the vicinity of 
$J_d/J_{\perp} \approx 0.5$ the system is made of weakly-coupled chains 
even when $J_{\perp}$ and $J_d$ are not small. This finding of the TSDMRG 
was checked using ED on small systems \cite{alvarez}. More careful 
simulations at the vicinity of the point $J_d/J_{\perp}=0.5$ revealed that 
the first neighbors interchain correlation, i.e. the transverse bond strength
 is equal to zero at the maximally frustrated point. The non-zero correlations,
 starting from the second neighbor decay exponentially. These results lead us 
 to conclude that the maximally frustrated point is a quantum critical point
(QCP) between the two  magnetic states (the second magnetic state is stable 
when $J_d > 0.5 J_{\perp}$). A possible argument against this conclusion
is  that at the maximally frustrated point, the system could be instable 
against higher orders terms such as a ring exchange term \cite{starykh}. 
In the case of a two-leg ladder this term seems to lead to a dimerized state. 
There are however some strong indications that the dimerized state does 
not exist in this model as discussed in Ref.\cite{moukouri-TSDMRG3}. The 
mechanism to avoid frustration is  to divide the system into chains in 
which the frustrated bonds are severed. We will now study the extension of 
this mechanism to $S=1$. We perform the same analysis as for $S=\frac{1}{2}$ 
for $L=6$, $8$, $10$ and $12$.  For the first three values of $L$, ED 
is performed to obtain the $m_2$ lowest eigenvalues and the corresponding 
eigenstates of the  chain. For $L=12$ the DMRG was used, we kept $m_1=243$ 
states, i.e, one DMRG iteration was done from the $L=10$ exact result. The 
truncation  error was about $7 \times 10^{-7}$. The truncation error is
relatively large because we used periodic boundary conditions.  
These $m_2$ low-lying states were  then used  to generate the 2D lattices, 
i.e, $6 \times 7$, $8 \times 9$, $10 \times 11$ and $12 \times 13$ 
respectively. 

\subsection{Ground-state energies}

\begin{figure}
\includegraphics[width=3. in, height=2. in]{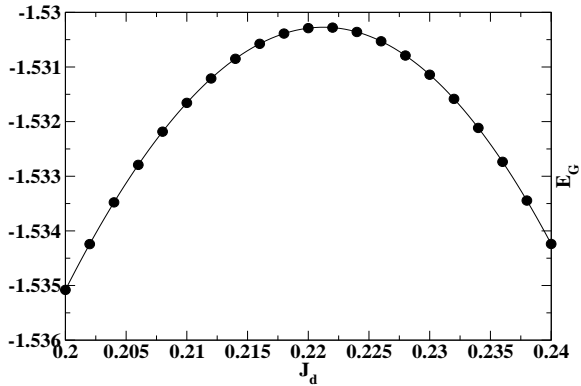}
\caption{Ground-state energy per site as function of $J_d$ for $J_{\perp}=0.4$, 
$S=1$ and $L=12$.}
\vspace{0.5cm}
\label{gs}
\end{figure}

\begin{figure}
\includegraphics[width=3. in, height=2. in]{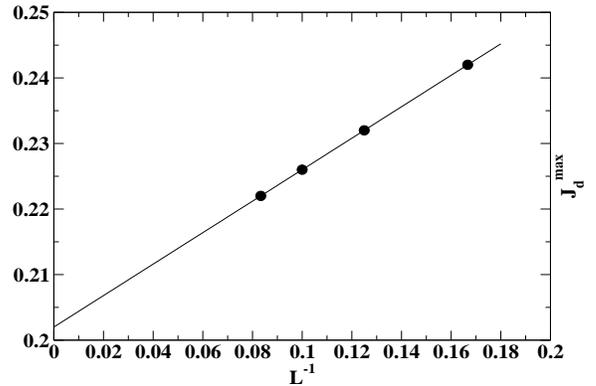}
\caption{Maxima of the ground-state energy as function of $L$ for
$J_{\perp}=0.4$, $S=1$.}
\vspace{0.5cm}
\label{maxgs}
\end{figure}

The curve of $E_G(J_d)$ for $S=1$, shown in Fig.(\ref{gs}) is  similar to 
that of $S=\frac{1}{2}$. Starting from $J_d=0$, $E_G$ increases until it 
reaches a maximum at $J_d^{max}$. It then decreases when $J_d$ is further 
increased. The position of the maximum depends slightly on $L$ and seems 
to converge to $0.5 J_{\perp}$ in the thermodynamic limit. $E_G(J_d^{max})$
is very close to $1.53$, the energy of decoupled chains, but always remains 
slightly lower. 
Thus as for $S=\frac{1}{2}$ the chains are very weakly bound, even though
the bare interactions ($J_{\perp}=0.4$ and $J_d=0.22$) are not small.
$J_d^{max}$ depends on $L$ as shown in Fig.(\ref{maxgs}) and as in the case 
of $S=\frac{1}{2}$, it extrapolates to $0.5 J_{\perp}$ in the 
thermodynamic limit.   

\begin{figure}
\includegraphics[width=3. in, height=2. in]{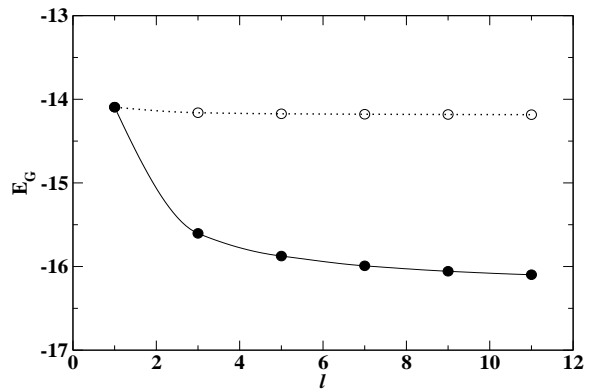}
\caption{Ground-state energy per chain as function of the chain number
for $J_{\perp}=0.4$, $S=1$, $L=10$ for $J_d=0$ (filled circles),
 and $J_d=J_d^{max}$ (open circles).}
\vspace{0.5cm}
\label{gsl}
\end{figure}

$E_G(l)$ shown in Fig.(\ref{gsl}), where $l$ is the number of chains is 
dramatically different when $J_d$ is far or close to $J_d^{max}$. Far 
from $J_d^{max}$, one of the two magnetic phases is highly favored. 
Starting from an isolated chain, magnetic energy can be gained by 
increasing $l$, leading to the ordered state. The situation
is different when $J_d=J_d^{max}$; neither of the magnetic states is 
favored. At this point, magnetic energy cannot be gained and $E_G(l)$ is
nearly independent of $l$, the system remains disordered as we will see
below from the analysis of the correlation functions.

\subsection{First neighbor correlation}

\begin{figure}
\includegraphics[width=3. in, height=2. in]{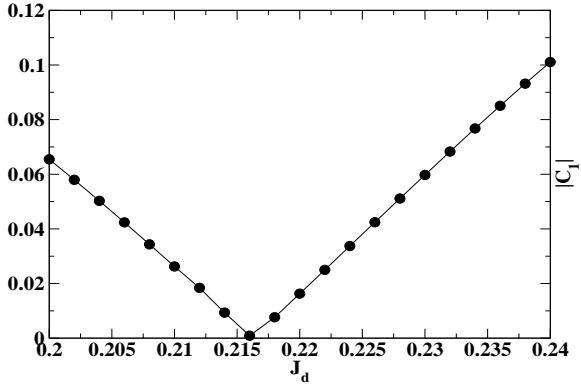}
\caption{First neighbor spin-spin correlation as function of $J_d$ for
$J_{\perp}=0.4$, $S=1$, $L=12$.}
\vspace{0.5cm}
\label{cor1}
\end{figure}

\begin{figure}
\includegraphics[width=3. in, height=2. in]{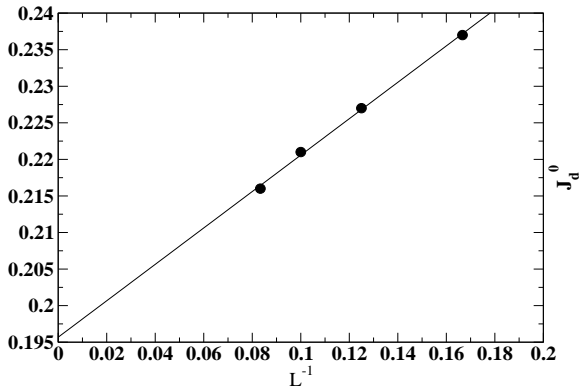}
\caption{Minima of the first neighbors spin-spin correlation as function of
$L$ for $J_{\perp}=0.4$, $S=1$.}
\vspace{0.5cm}
\label{mincor1}
\end{figure}

The transverse first neighbor spin-spin correlation taken in the middle
of the lattice 

\begin{equation}
C_1=\langle S_{L/2,L/2+1}S_{L/2,L/2+2} \rangle, 
\end{equation}

\noindent shown in 
Fig.(\ref{cor1}) is also reminiscent
of the $S=\frac{1}{2}$ case. $C_1$ vanishes linearly at $J_d=J_d^0$. 
$J_d^0$ is slightly different from $J_d^{max}$.  This small difference is 
due to numerical error; this conclusion is  supported by the 
more accurate results obtained for small $S=\frac{1}{2}$ systems in 
Ref.(\cite{moukouri-TSDMRG3}) where
$J_d^0$ and $J_d^{max}$ are equal. The extrapolated $J_d^0$ 
(Fig.(\ref{mincor1}) as $L \rightarrow \infty$ is also in the vicinity
of $0.5 J_{\perp}$.    

\subsection{Long-distance correlations}

\begin{figure}
\includegraphics[width=3. in, height=2. in]{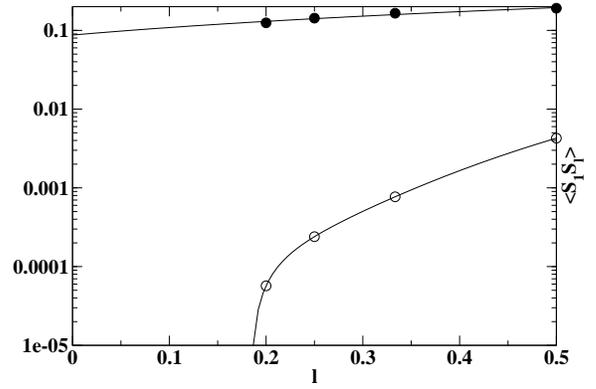}
\caption{Transverse spin-spin correlation as function of the distance
for $L=10$, $S=1$, $J_{\perp}=0.4$, $J_d=0$ (filled circles) and 
$J_d=J_d^{max}$ (open circles).}
\vspace{0.5cm}
\label{corl}
\end{figure}

\begin{figure}
\includegraphics[width=3. in, height=2. in]{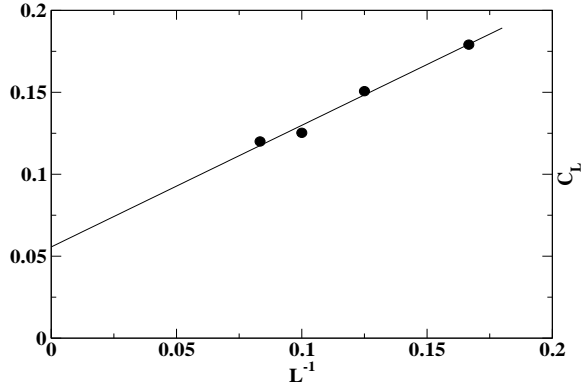}
\caption{Center-to-end spin-spin correlation as function of $L$ for
$S=1$, $J_{\perp}=0.4$ and $J_d=0$.}
\vspace{0.5cm}
\label{core}
\end{figure}

The transverse spin-spin correlation function, 

\begin{equation}
C_l=\langle S_{L/2,L/2+1}S_{L/2,L/2+l} \rangle
\end{equation}

\noindent is shown in 
Fig.(\ref{corl}) for a $L=10$ system and $J_{\perp}=0.4$. 
When $J_d=0$, $C_l$ extrapolates
to a finite value in the thermodynamic limit. This result is important 
because it shows the non-perturbative nature of the TSDMRG. Starting from
an isolated chain which is disordered, the TSDMRG can reach the ordered
phase. The ordered phase can be easily reached for larger spin than
for $S=\frac{1}{2}$ where quantum fluctuations are more important.
The extrapolation of $C_l$ for $S=\frac{1}{2}$ leads to a small negative 
value as shown in Fig.(\ref{corlS}) below. In this case, the order 
parameter is too small to be obtained from an extrapolation from relatively 
small systems; it is necessary to go to larger systems as those studied
in Ref.(\cite{moukouri-TSDMRG2}) in order to extrapolate to the correct
thermodynamic limit. The extrapolated value of $C_l$ does not however
lead to the correct value of the magnetization since it is obtained 
from a system with a fixed $L$. A better estimation is given by 
the finite size analysis of the end-to-center spin-spin correlation

\begin{equation} 
C_L=\langle S_{L/2,L/2+1}S_{L/2,L+1} \rangle
\end{equation}

\noindent shown in Fig.(\ref{core}). $C_L$ for $L \rightarrow \infty$ is 
roughly $0.06$ consistent again with the existence of the long-range order.

In the vicinity of $J_d^{max}$, $C_l$ decays exponentially
as seen in Fig.(\ref{corl}). For $l=5$, his value is already four
order of magnitude smaller than in the case $J_d=0$. This is consistent
the nearly disconnected chain behavior observed for $E_G$ at this point.

\subsection{spin gaps}

\begin{figure}
\includegraphics[width=3. in, height=2. in]{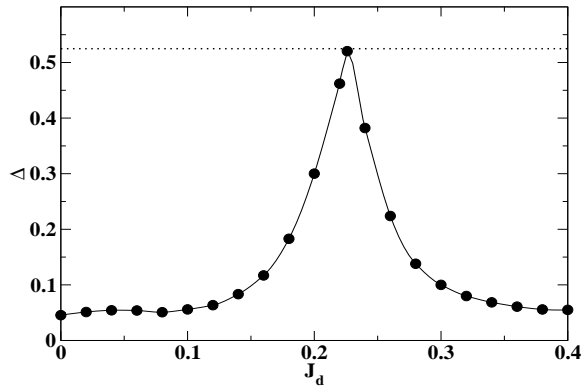}
\caption{Gap as function of $J_d$ for $L=10$, $S=1$; the dotted line
represent the single chain gap.}
\vspace{0.5cm}
\label{gap}
\end{figure}

\begin{figure}
\includegraphics[width=3. in, height=2. in]{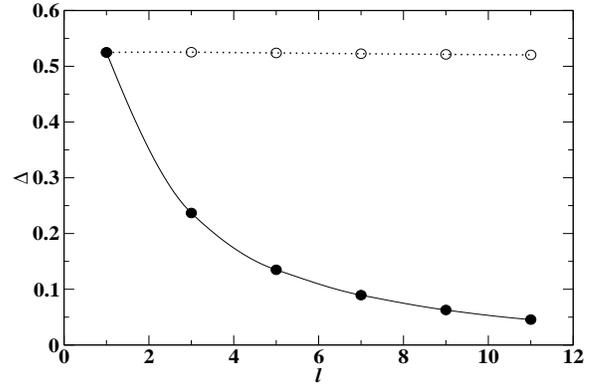}
\caption{Gap as function of the number of chains for $L=10$, $S=1$,
$J_{\perp}=0.4$, $J_d=0$ (filled circles) and $J_d=J_d^{max}$ (open circles).}
\vspace{0.5cm}
\label{gapl}
\end{figure}

The variation of the spin gap $\Delta$ with $l$ (for a fixed $L$), $L$,
and $J_d$ the  are also consistent with 
the above findings. $\Delta(J_d)$ for $L=10$ system is shown in 
Fig.(\ref{gap}). $\Delta(J_d=0)$ is about $0.05$ (in this regime the gap 
is zero in the thermodynamic limit as we will see below); $\Delta$ remains 
relatively flat as $J_d$ is increased until it reaches the vicinity of 
$J_d^{max}$. Near $J_d^{max}$, $\Delta(J_d)$ first sharply increases and 
reaches the finite size gap of an isolated chain. As $J_d > J_d^{max}$, 
$\Delta(J_d)$ first sharply decreases and then becomes nearly constant at 
about $0.05$. $\Delta(l)$ (Fig.(\ref{gapl})) is reminiscent of $E_G(l)$; 
in the unfrustrated case, the chains are effectively coupled. $\Delta(l)$ 
rapidly decreases from about $0.53$ to $0.05$ as $l$ is varied from $1$ to 
$11$. At $J_d=J_d^{max}$ however, $\Delta(l)$ is nearly independent of $l$.

\begin{figure}
\includegraphics[width=3. in, height=2. in]{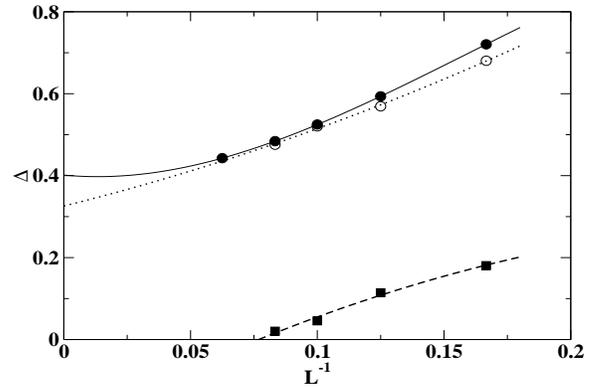}
\caption{Gap as function of $L$ for $S=1$, $J_{\perp}=0$ and
$J_d=0$ (single chain, filled circles), $J_{\perp}=0.4$, $J_d=0$
(filled squares) and $J_d=J_d^{max}$ (open circles).}
\vspace{0.5cm}
\label{gapL}
\end{figure}

The analysis $C_L$ shows that for $J_d=0$, the system is ordered. We thus
expect that $\Delta(L) \rightarrow 0$ as $L \rightarrow \infty$. This is seen
in Fig.(\ref{gapL}) where $\Delta(L)$ is shown for $L=6,8,10$ and $12$ 
systems. The decay faster than $\frac{1}{L}$,  the extrapolation to
leads to a negative value. At $J_d=J_d^{max}$ on the order hand, $\Delta(L)$
remains close to that of an isolated chain in all case. The two
functions are finite in the thermodynamic limit. This result show the
dramatic difference between $S=\frac{1}{2}$ and $S=1$ systems. For 
$S=\frac{1}{2}$, an equivalent plot lead to a zero gap \cite{moukouri-TSDMRG3}
at the maximally frustrated point. The extrapolated
value for $J_d=0$, $\Delta=0.4015$ agrees well with the current best
estimate of the Haldane gap $\Delta_H=0.4107$. The difference is due to
the relatively short chains, up to $L=16$,  that were used for the
extrapolation not to the DMRG that yielded highly accurate results for each
size studied. Noting that the spin-spin correlations in the transverse
direction have a very short range, the difference between the extrapolated 
values of $J_d=0$ and $J_d=J_d^{max}$ appears to be relatively large. 
We believe that this difference could be inferred from the fact that the
extrapolation from $2D$ systems were done with lattice sizes up to $L=12$
only.      

\subsection{Spin-spin correlations on large systems}

So far, in the study of $S=1$ systems, we have fixed $J_{\perp}=0.4$.
This choice was motivated by the presence of $\Delta_H=0.4107$ in an
isolated chain. We initialy felt that it was necessary to choose a large
enough $J_{\perp}$ so that the chains will effectively be coupled when
the perturbation is turned on and this will lead to sizable correlation in the thermodynamic limit. But, this choice limited us to relatively
small lattices, $L \alt 12$. This is because when $J_{\perp}$ is large,
the condition $\delta E/J_{\perp} \gg 1$ is hard to fulfill for larger $L$.
For instance for $L=16$, $\delta E/J_{\perp}=5.65$ for $m_2=64$,
 this prevented us to
study $L=16$ lattices. But in the course of this work, we find that even
smaller values of $J_{\perp}$ can lead to detectable values of in the
unfrustrated regime $C_l$ as $l \rightarrow \infty$. For smaller $J_{\perp}$,
we can actually reach larger $L$. We wish to present in this part our
results for $J_{\perp}=0.2$ and $L=24$. These results will add strength
to those of $J_{\perp}=0.4$ presented above.

\begin{figure}
\includegraphics[width=3. in, height=2. in]{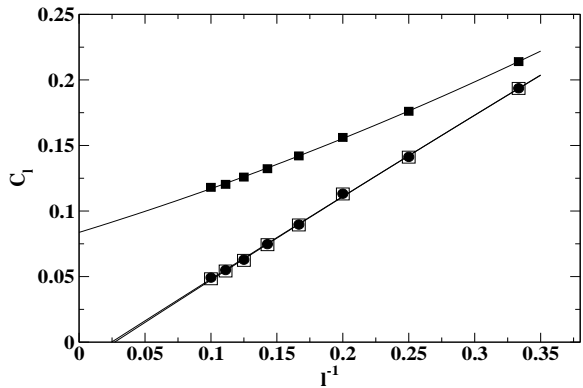}
\caption{Longitudinal spin-spin correlation as function of the distance
for $L=24$, $S=1$, $J_{\perp}=0$, $J_d=0$ (filled circles), $J_{\perp}=0.2$,
$J_d=0$ (filled squares), $J_d=0.2$, $J_d=0.102$ (open squares).}
\vspace{0.5cm}
\label{corl24}
\end{figure}

\begin{figure}
\includegraphics[width=3. in, height=2. in]{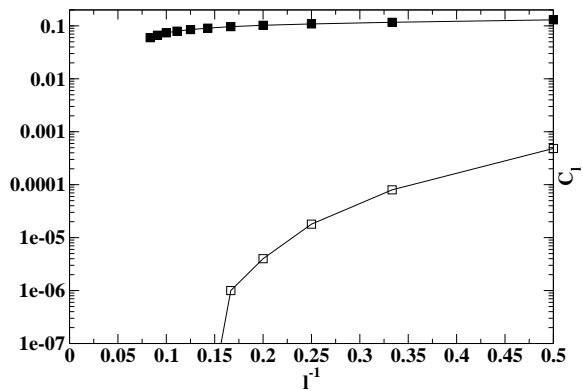}
\caption{Transverse spin-spin correlation as function of the distance
for $L=24$, $S=1$, $J_{\perp}=0$, $J_d=0$ (filled circles), $J_{\perp}=0.2$,
$J_d=0$ (filled squares), $J_d=0.2$, $J_d=0.102$ (open squares).}
\vspace{0.5cm}
\label{corlt24}
\end{figure}

In Fig.(\ref{corl24}), we show the longitudinal correlation,

\begin{equation}
{\bar C}_l=\langle S_{L/2+1,L/2+1}S_{L/2+l,L/2+1} \rangle,
\end{equation}

\noindent taken in the middle chain. For $J_{\perp}=0.2$ and $J_d=0$,
${\bar C}_l$ clearly extrapolate to a finite value as expected. But
for $J_{\perp}=0.2$ and $J_d=0.102$, ${\bar C}_l$ is nearly identical
to spin-spin correlation on an isolated chain. For the transverse
correlation $C_l$ shown in Fig.(\ref{corlt24}), we see again the 
dramatic difference between the
unfrustrated and highly frustrated cases. In the first case, $C_l$ goes 
to a finite value when $l \rightarrow \infty$. But for the highly
frustrated case $C_l$ decays exponentially.

\begin{figure}
\includegraphics[width=3. in, height=2. in]{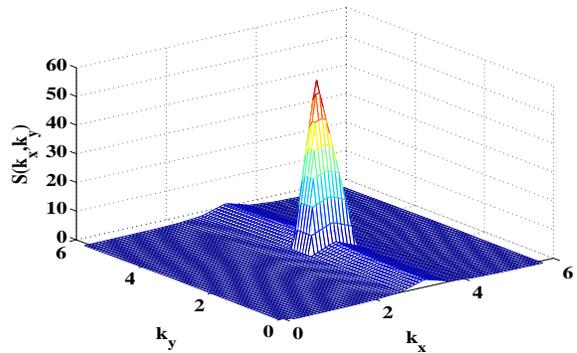}
\caption{Magnetic structure factor 
for $L=24$, $S=1$, $J_{\perp}=0.2$, $J_d=0$.}
\vspace{0.5cm}
\label{sfac124}
\end{figure}

\begin{figure}
\includegraphics[width=3. in, height=2. in]{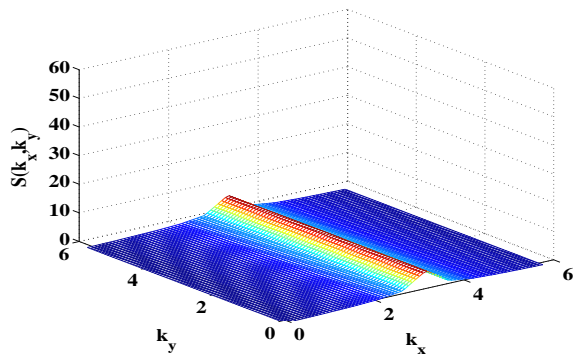}
\caption{Magnetic structure factor
for $L=24$, $S=1$, $J_{\perp}=0.2$, $J_d=0.102$.}
\vspace{0.5cm}
\label{sfac224}
\end{figure}

Another picture of this dramatic difference is given by the magnetic
structure factor $S(Q=(q_x,q_y))$ shown in Fig.(\ref{sfac124},\ref{sfac224}).
 In the magnetic phase, $S(Q)$ is dominated
by a sharp peak at $Q=(\pi,\pi)$ indicative of N\'eel order. 
In the disordered phase, $S(Q)$ is nearly flat except for a small range 
along $Q=(\pi,q_y)$ which retains the signature of short-range in chain 
correlations.

\subsection{Conclusion}

In this section, we have presented comprehensive results on
$S=1$ coupled chains. These results show some analogy with those of
$S=\frac{1}{2}$ published in Ref.(\cite{moukouri-TSDMRG3}). Starting
from the unfrustrated system for which $J_d=0$, the ground state for
$S=1$ is ordered, as expected from the DLS theorem. While for
$S=\frac{1}{2}$ it was necessary to simulate lattices of up to $L=64$
\cite{moukouri-TSDMRG2} in order to see the extrapolation of $C_l$ to
a finite value, for $S=1$, relatively short sizes ($L=12$) were enough.
This is indeed because quantum fluctuations are less important in a $S=1$
system, i.e., the order parameter is larger. This enables it to be
computed more easier. By comparison, the same extrapolation done for 
$S=\frac{1}{2}$ will lead to a negative value.

When the frustration $J_d$ is turned on and reaches the value $J_d^{max}$,
a point where $E_G$ is maximum, $C_l$ decays exponentially, $\Delta$
takes a value very close to that of a pure 1D system. These results
imply that at $J_d^{max}$, the transverse interactions are irrelevant.
For $S=\frac{1}{2}$, we identified this state as spin version of an SLL.
In that case, this sliding phase will probably be confined at the critical
point where the two competing magnetic states $Q=(\pi,\pi)$ and $Q=(\pi,0)$ 
neutralizes each other. However, we cannot completely rule out a small
finite extension of the sliding phase or even totally exclude the emergence
of a relevant interaction at lower energies which eventually drives the
system to a dimerized phase \cite{starykh}. It is obvious that, though
both $S=\frac{1}{2}$ and $S=1$ systems are made of nearly disconnected 
chains at $J_d^{max}$, the conclusions must be different because of the
presence of the Haldane gap $\Delta_H$ in the $S=1$ chain. The 
existence of $\Delta_H$ restricts the possible phases that may arise in
the vicinity of $J_d^{max}$. The first crucial difference is that the
disordered state that exist in an $S=1$ system has a gap in its excitation
spectrum as seen in Fig.(\ref{gapL}); any eventual residual interaction
will be wiped out by this gap which means the emergence of new phases at
low energies  is not favorable for an $S=1$ system. This disordered phase
has probably a finite extension which is roughly $\propto \Delta_H$.

\section{Results for $S=\frac{1}{2}$ to $S=4$}
\label{spinS}

In the study of various $S$, we will not do the same extensive calculation
seen in the preceeding section with $S=1$. We will simply fix $L$ and 
analyze the behavior of the system as function of $J_d$ and $l$. As we 
will see below, quantum fluctuations are small for $S >1$, the study
of relatively small systems is enough to get the correct picture in the
thermodynamic limit. We studied a lattice with $L \times (L+1)=10 \times 11$ 
for $S=\frac{1}{2}$ to $S=4$. $J_{\perp}$ was set to $0.4$ so that it is 
larger than the finite size gap in spin half-integer systems and larger or 
close to the Haldane gaps in spin integer systems. $J_d$ is varied from 
$0$ to $0.4$.

\subsection{Ground-state energies}

\begin{figure}
\includegraphics[width=3. in, height=2. in]{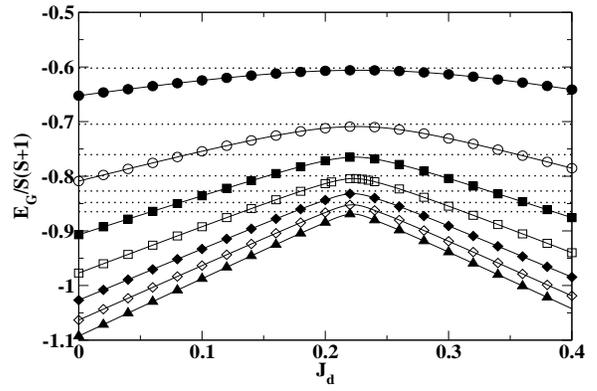}
\caption{ Ground state energy as function of $J_d$ and 
$S=\frac{1}{2}$ (filled circles), $S=1$ (open circles),
$S=\frac{3}{2}$ (filled squares), $S=2$ (open squares),
$S=\frac{5}{2}$ (filled diamonds), $S=3$ (open diamonds),
$S=\frac{7}{2}$, (filled triangles).}
\vspace{0.5cm}
\label{gsS}
\end{figure}

The ground-state energy is shown in Fig.(\ref{gsS}) for $S=\frac{1}{2}$
to $\frac{7}{2}$. Simulations were also done for $S=4$ but they did not
converge for certain values of $J_d$. We infer this failure to the
large degeneracy of the renormalized single chain Hamiltonian $h_l$ for 
large $S$. At $J_d=0$, the curves approach the classical value 
$E_G/S^2=-1.4$ quite rapidly. For $S=\frac{7}{2}$, we find $E_G/S(S=1)=-1.093$.
But if we use the $1/S^2$ normalization, we get  $E_G/S^2=-1.405$. Hence
if both finite size effects and the correct normalization are taken into
account, $S=\frac{7}{2}$ is already in the classical limit.

\begin{figure}
\includegraphics[width=3. in, height=2. in]{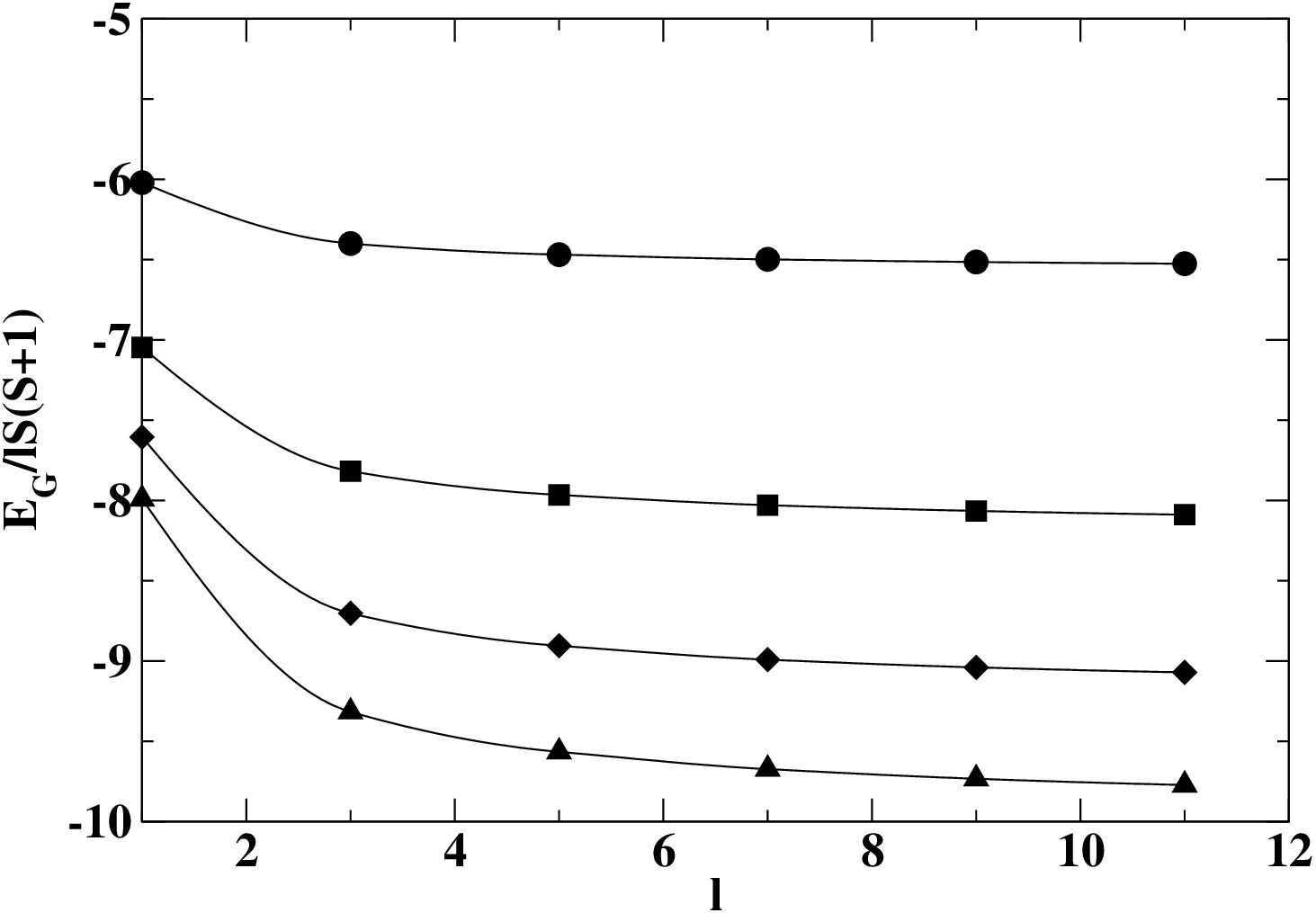}
\caption{Ground state energy at $J_d=0$ as function of $l$ and $S=\frac{1}{2}$
(circles), $S=1$ (squares), $S=\frac{3}{2}$ (diamonds), $S=2$ (triangles)}
\vspace{0.5cm}
\label{gSl1}
\end{figure}

\begin{figure}
\includegraphics[width=3. in, height=2. in]{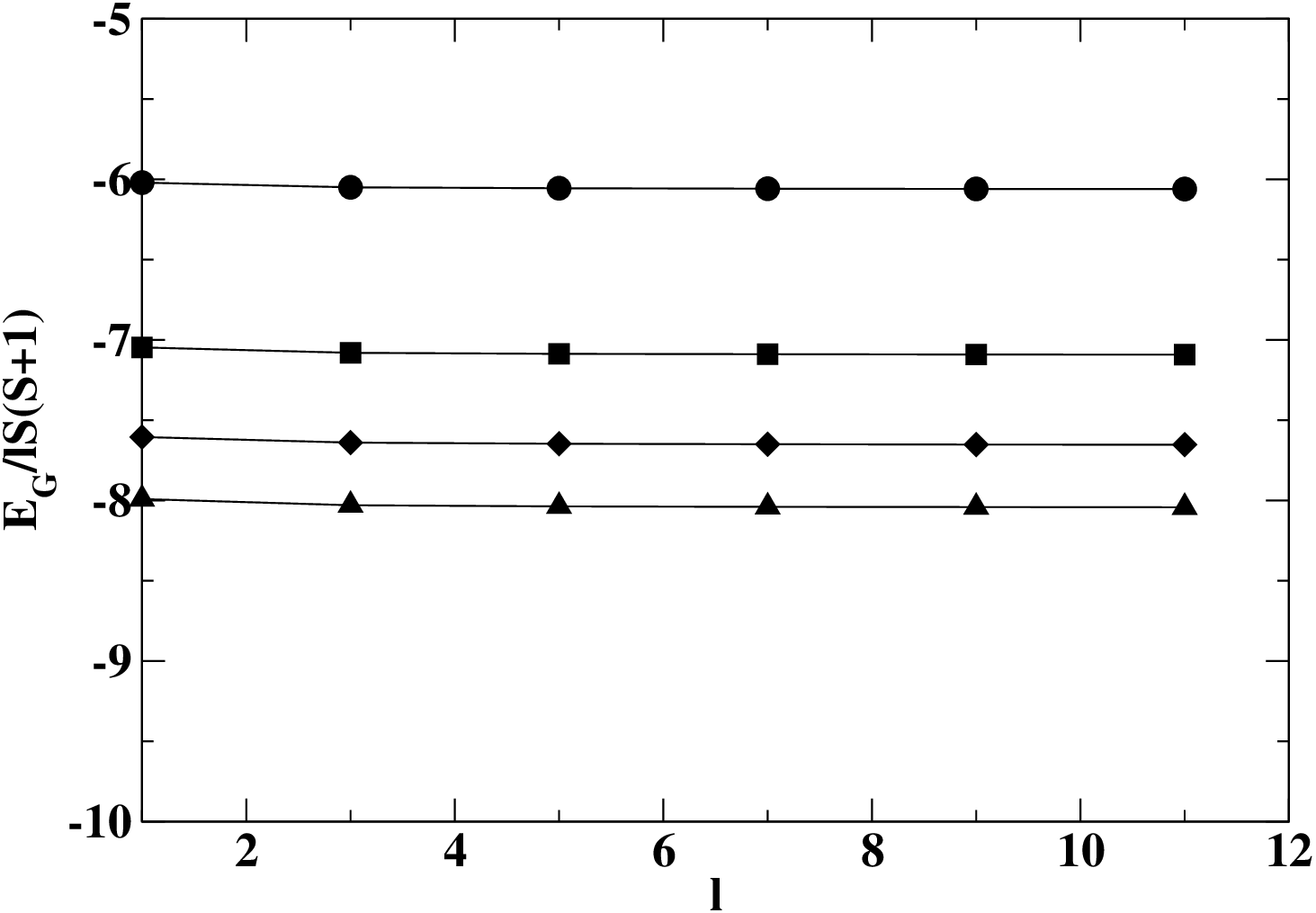}
\caption{Ground state energy at $J_d=0$ as function of $l$ and $S=\frac{1}{2}$
(circles), $S=1$ (squares), $S=\frac{3}{2}$ (diamonds), $S=2$ (triangles)}
\vspace{0.5cm}
\label{gSl2}
\end{figure}

Larger spin systems are found to present the same features displayed by spin 
$1/2$ systems. The ground-state energy, $E_G$, shown in Fig.(\ref{gsS}) 
increases as $J_d$ increases until the maximally frustrated point where $E_G$ 
of the two-dimensional system becomes very close to that of disconnected 
chains. From this point it decreases when $J_d$ is further increased. 
This may be interpreted as follows: starting from the N\'eel state with 
$Q=(\pi,\pi)$ for $J_d=0$, the system tends to lose energy under the action 
of $J_d$ which progressively destroys the N\'eel order until the maximally 
frustrated point. Beyond this point, $J_d$ becomes dominant and the systems 
enters the N\'eel $Q=(\pi,0)$ phase. The position of this maximum decreases 
slowly with increasing $S$. This indicates that in addition to the effect 
of OBC that shifts $J_d^{max}$ towards higher values, there are intrinsic 
finite size effects. All systems evolve regularly towards the 
$S \rightarrow \infty$ limit. 

The curve of $E_G$ appears to change structure as $S$ increases. At low
$S$, a well-rounded maximum is observed. We were able to fit all the
points of the curve to a quadratic function. But for large $S$, this
became impossible. The maximum has nearly become a cusp as for 
$S \rightarrow \infty$. This cusp is at the intersection of two straight
lines $E_{G_1}=-1-J_{\perp}+2J_d$ and $E_{G_2}=-1+J_{\perp}-2J_d$ which
are the ground-state energies, respectively, below and above the 
transition point $J_d=0.5 J_{\perp}$. 

$E_G(l)$ shown in Fig.(\ref{gSl1},\ref{gSl2}) displays the features seen for $S=1$.
It decreases when $l$ increases in the weak frustration regime. It remains
nearly constant in the vicinity of the maximally frustrated point. 
\subsection{First neighbor correlation}

\begin{figure}
\includegraphics[width=3. in, height=2. in]{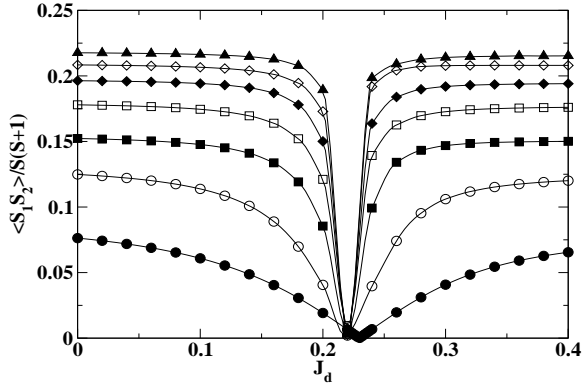}
\caption{First neighbor correlation as function of $J_d$ and $S=\frac{1}{2}$
(circles), $S=1$ (squares), $S=\frac{3}{2}$ (diamonds), $S=2$ (triangles)
$S=\frac{5}{2}$ (filled diamonds), $S=3$ (open diamonds),
$S=\frac{7}{2}$, (filled triangles).}
\vspace{0.5cm}
\label{cor1S}
\end{figure}

The tendency to the severing of the chains is more clearly seen in the 
transverse bond strength

\begin{equation}
C_1=\langle S_{5,6}^zS_{5,7}^z \rangle,
\end{equation}

\noindent shown in Fig.(\ref{cor1S}). In all cases, $C_1$ decreases 
from its value at $J_d=0$ and 
seems to vanish at $J_d^0$ (we were able in all cases to reach values 
of $C_1$ which are equal or less than the numerical accuracy of in our 
simulations). From this point it increases. There is a small difference 
between the position of $J_d^0$ for different values of $S$ as found 
for $E_G$. The curves of $C_1$ suggest that for all $S$, the mechanism
to avoid frustration is identical: the systems relaxe to nearly disconnected
chains. These curves also show the influence of quantum fluctuations for
small $S$. This is seen in the decay of $C_1$ as soon as $J_d \neq 0$. For
larger $S$, $C_1$ remains nearly constant until $J_d \approx J_d^0$.

\subsection{Long-distance correlations}

\begin{figure}
\includegraphics[width=3. in, height=2. in]{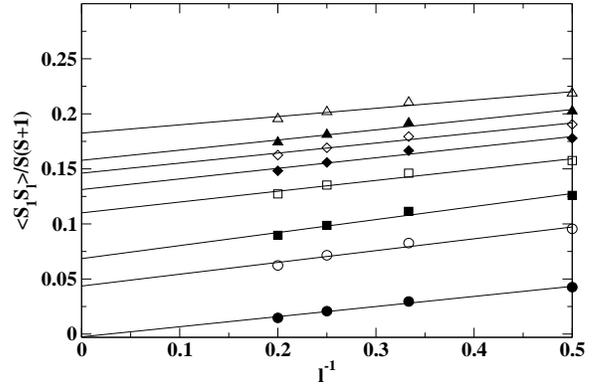}
\caption{ Spin-spin correlation at $J_d=0$ as function of $l$ and
$S=\frac{1}{2}$ (filled circles), $S=1$ (open circles),
$S=\frac{3}{2}$ (filled squares), $S=2$ (open squares),
$S=\frac{5}{2}$ (filled diamonds), $S=3$ (open diamonds),
$S=\frac{7}{2}$, (filled triangles), $S=4$ (open triangles).}
\vspace{0.5cm}
\label{corlS}
\end{figure}

Since our starting point for 2D systems is disconnected chains, it is 
important to show, as for the spin $1/2$ case studied previously, that 
the TSDMRG  is able to reach the ordered phase. One possible way to look 
at the appearance of the ordered state is to look at the decay of the 
transverse correlation function, 

\begin{equation}
C_l=\langle S_{5,6}^zS_{5,5+l}^z\rangle,
\end{equation}

\noindent in the N\'eel 
phase for $J_d=0$.  We found that for all 
values of $S$ except $S=\frac{1}{2}$, as shown in Fig.(\ref{corlS}), the 
transverse correlation function extrapolate to finite values. $C_l$ 
extrapolates to a negative value for $S=\frac{1}{2}$. In that case, 
quantum fluctuations are so strong that it is necessary to go to larger 
values of $L$ as done in Ref.(\cite{moukouri-TSDMRG2}).

At the maximally frustrated point, we also observe similar 
exponential decay of $C_l$ as for 
$S=\frac{1}{2}$. This decay is less faster with increasing $S$. Indeed 
in the limit $S \rightarrow \infty$, the transition is of first order. 
The chains are disconnected in this classical limit and $C_l$ is exactly 
equal to zero. However for any small deviation from the transition point, 
the system falls in one of the ordered states. This point is virtually 
impossible to find exactly numerically. However, for smaller values of $S$ 
the critical region is larger and even if we miss the exact transition point,
 this  behavior will nevertheless be observed as far as we close 
enough to the QCP.

\begin{figure}
\includegraphics[width=3. in, height=2. in]{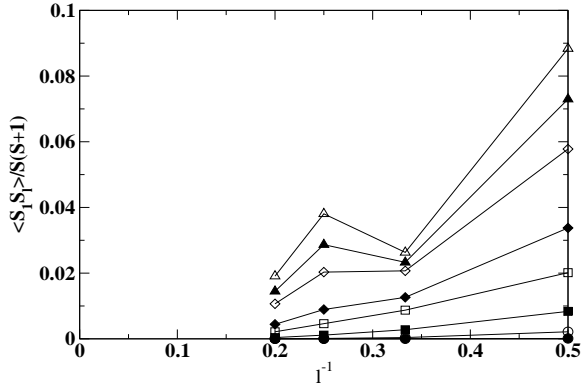}
\caption{Transverse spin-spin correlation at $J_d=0.22$ as function of $l$ 
and $S=\frac{1}{2}$ (circles), $S=1$ (squares), $S=\frac{3}{2}$ (diamonds), 
$S=2$ (triangles), $S=\frac{5}{2}$ (filled diamonds), $S=3$ (open diamonds),
$S=\frac{7}{2}$, (filled triangles), $S=4$ (open triangles).}
\vspace{0.5cm}
\label{corlS2}
\end{figure}

\subsection{Spin gaps}

\begin{figure}
\includegraphics[width=3. in, height=2. in]{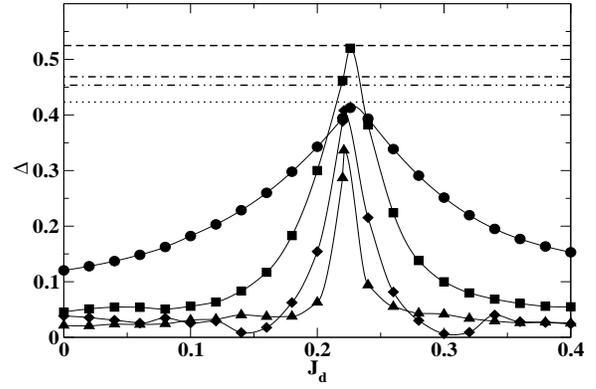}
\caption{Spin gap as function of $J_d$ and $S=\frac{1}{2}$
(circles), $S=1$ (squares), $S=\frac{3}{2}$ (diamonds), $S=2$ (triangles)}.
\vspace{0.5cm}
\label{gapS}
\end{figure}

The curves of $\Delta(J_d)$ in Fig.(\ref{gapS}) for different values 
of $S$ have typically
a peak in the at $J_d=J_d^{max}$. This peak is very narrow, except for
$S=\frac{1}{2}$ where  quantum fluctuation effects lead to a broader
peak. As expected from the behavior of $C_1$, this peak is sharper with
increasing $S$. $\Delta(J_d^{max})$ is nearly equal to the finite size gap 
of an isolated chain which is represented by a flat line in each case.
We were unable to reach the 1D gap for $S > \frac{3}{2}$. This is probably
due to the narrowness of the critical region which makes it difficult 
to see the nearly disconnected chain regime. One can easily fall in one
of the ordered regime leading to a relative slower decay of $C_l$ which
manifests itself to a smaller finite size gap.

\subsection{Conclusion}

In this section, we presented results for $L=10$ and $J_{\perp}=0.4$ with
$S$ varying from $\frac{1}{2}$ to $4$. In agreement with the DLS theorem,
we found long-range order for all $S$ greater or equal to $1$ in the
unfrustrated case. As $J_d$ is turned on, the long-range order is 
destroyed. An interesting question is the nature of this disordered state
as function of $S$. Before addressing this question, we will first review 
how frustration works in 1D \cite{schollwock}.  

For the frustrated $S=\frac{1}{2}$ chain, there is a 
transition to a dimerized phase at $J_{2c}=0.242 J_1$, where $J_{2c}$ is
the value of next-nearest neighbor coupling at the critical point.
At this point a gap opens exponentially and grows with $J_2$. At 
$J_2=0.5 J_1$, the system is perfectly dimerized and shows incommensurate
correlation above this point (disordered point). At about $J_2=0.52 J_1$
a two-peak structure appears in the structure factor (Lifshitz point).
 DMRG simulations for the $S=\frac{3}{2}$ chain shows a similar behavior.
This suggests that it is generic to half-integer spin systems.

Integer spin chains are already gapped in the absence of the frustration
term $J_2$. For $S=1$, the  transition to a dimerized phase is absent, 
but numerical simulations show the presence of the disordered and the 
Lifshitz points. In addition there is a first order transition at
$J_2=0.75 J_1$ from a phase with a single string order to a phase with a 
double string order. At this point the chain splits into two chains. 
These special points are also observed in the 
$S=2$ chain except that the order parameter for the first order transition
is still unknown. 

 The results presented in the preceding sections show that the mechanism 
to ease frustration works differently in 2D systems. This mechanism is
 the same for all values of $S$. The system spontaneously severs the 
frustrated bond at the maximally frustrated point. The similarity for 
all $S$ of this mechanism stems from the fact that if the transverse 
coupling is large enough, all the 2D systems are ordered for half-integer 
as well as for odd integer systems. Frustration is a 
competition between two magnetic ground phases and we have shown that for 
coupled chain systems, the best way to avoid frustration is to relax into 
nearly independent chains. It is clear that such a mechanism will 
be independent on the value of the spin as found in our numerical study. 
The consequences are nevertheless different for half-odd integer and for 
integer spin systems.

 For all spin half-odd integer systems, like for the 
spin $1/2$ studied more extensively in Ref.(\cite{moukouri-TSDMRG3}), 
there is a second order 
phase transition between the two magnetic states at $J_d=J_d^{max}$. At the 
critical point, the system is disordered. The transverse correlation decay 
exponentially while at long distances, the longitudinal one behave like those
 of independent chains. Hence at the critical point, the spin rotational 
symmetry of the Hamiltonian is restored. As for spin $1/2$, there might be 
a residual interaction which can drive the system eventually to a SP phase. 
But,  previous numerical studies on 2D systems \cite{moukouri-TSDMRG3} and
on three-leg ladders point to an absence of a dimerized phase in this region
for $S=\frac{1}{2}$. This is expected to be valid for all half-odd $S$. 
We would like to stress that dimerization is not the driving mechanism
mechanism in the formation of the disordered state. We are indeed aware
of earlier ED results \cite{dagotto} in which an enhancement of the
SP susceptibility was observed in the regime $J_2 \approx J_1/2$. We
believe in the light of our results that this merely the consequence
of the severing of the chains in one of the two directions of the
square lattice. The SP signal is expected to be larger in 1D where it
has a power law decay than in 2D when the spins are locked into 
N\'eel order in the unfrustrated regime.

For integer spins, there is an intermediate phase between the 
two magnetic states. When $|J_{\perp}-2J_d| \ll \Delta$, where $\Delta$ is 
the single chain spin gap, the transverse coupling are irrelevant. The
maximally frustrated point is the equivalent of the disordered point
seen in 1D. In this 
regime of couplings, the system is an assembly of nearly decoupled chains. 
In the 
case of integer spins, even if there is a residual interaction at the 
maximally frustrated point, this interaction is necessary irrelevant 
because of the presence of $\Delta$. Integer spin systems are thus 
radically different from half-odd integer systems.

\section{Conclusion}
\label{conclusions}

In this paper, we used the TSDMRG to study coupled spin chains with $S$
varying from $\frac{1}{2}$ to $4$. This study illustrates the power of
the TSDMRG method, where using a modest computer effort 
we were able to study the unfrustrated regime and find long-range magnetic 
order, in agreement with the DLS theorem and Monte Carlo studies.
 We obtained  good accuracy in the highly frustrated regime
of the model.  The study of this region has so far resisted to other 
numerical methods.

We showed that in order to avoid frustration, all spin systems tend to 
sever the frustrated bonds. The severing of the transverse bonds is a 
large effect which is seen in various physical quantities. The strong
frustration regime is dominated by 1D physics, topological effects 
become important as predicted in Ref.(\cite{affleck1,haldane2,read}).
However, we did not find any qualitative difference between odd and even
integer spin systems as predicted in Ref.(\cite{haldane2, read}). It
could be due to the fact that in the highly frustrated regime the 2D systems 
tend to relax into nearly independent 1D sytems where tolological effects 
are identical for odd and even integer spins. It could also be related
to the anisotropy of the model studied.

\begin{acknowledgments}
The author wishes to thank K. L. Graham for reading the manuscript. 
This work was supported by the NSF Grant No. DMR-0426775.
\end{acknowledgments}

\end{document}